\documentclass[prl,twocolumn,superscriptaddress]{revtex4}
\usepackage{amssymb,amsmath,amsthm,graphicx}
\usepackage{latexsym}
\usepackage{bm}
\usepackage{cleveref}
\usepackage{tabularx}

\pdfoutput=1

\begin{document}

\title{New islands of stability with double-trace deformations}

\author{Ramon~Masachs}
\email{rmg1e15@soton.ac.uk}
\affiliation{STAG research centre and Mathematical Sciences, University of Southampton, Highfield Southampton SO17 1BJ, UK}

\author{Benson~Way}
\email{benson@phas.ubc.ca}
\affiliation{Department of Physics and Astronomy, University of British Columbia, 6224 Agricultural Road, Vancouver, B.C., V6T 1W9, Canada}

\begin{abstract}
We assess the role of a resonant spectrum in the AdS instability, and quantify the extent to which breaking the resonant spectrum of AdS can restore stability.  Specifically, we study non-collapsing `multi-oscillator' solutions in AdS under various boundary conditions that allow for both resonant and non-resonant spectra.  We find non-collapsing two mode, equal amplitude solutions in the non-resonant Robin case, and that these solutions vanish in the fully resonant Dirichlet case.  This is consistent with non-resonant stability, and with the idea that stable solutions in the Dirichlet case are all single-mode dominated. Surprisingly, when the boundary condition is Neumann, we find non-collapsing solutions arbitrarily close to AdS that are not single-mode dominated, despite the spectrum being fully resonant.
\end{abstract}

\maketitle

{\bf~Introduction --}  Because of the presence of a reflecting boundary, global anti-deSitter space (AdS) allows for arbitrarily small excitations to form black holes, which implies that AdS is nonlinearly unstable \cite{DafermosHolzegel2006,Dafermos2006}. The first evidence of this instability was presented in \cite{Bizon:2011gg} where the numerical evolution of a massless scalar field in AdS, initially in a Gaussian configuration, eventually forms an event horizon even for very small amplitudes.  The existence of arbitrarily small data that inevitably form black holes was proven rigorously for the spherically symmetric pressureless Einstein-massless Vlasov system in \cite{Moschidis:2017lcr,Moschidis:2017llu}.

However, it was realised that there are configurations that do not form black holes \cite{Dias:2012tq,Maliborski:2013jca,Buchel:2013uba,Balasubramanian:2014cja,Bizon:2014bya,Balasubramanian:2015uua,Dimitrakopoulos:2015pwa,Green:2015dsa}.  The set of such non-collapsing initial data is called the {\emph{islands of stability}}.  This naturally leads to the open question: what distinguishes collapsing data from non-collapsing data?  Through the AdS/CFT correspondence \cite{Maldacena:1997re,Gubser:1998bc,Witten:1998qj}, the answer to this question has important implications for the process of thermalization and equilibration in the dual field theory.

Some guidance can be provided by perturbation theory.  The linear spectrum of perturbations of AdS consists of normal modes with evenly-spaced frequencies.  When expanding a generic configuration to higher orders in perturbation theory, resonant self-interactions from this spectrum causes secular terms to appear that predict the breakdown of perturbation theory.  Moreover, the timescale for perturbative breakdown matches the time scale for horizon formation.  However, for the special cases of single-mode data (and only these cases), these secular terms can be removed by shifts in the frequency, and perturbation theory can continue to arbitrarily high orders.  The emergent picture, as supported by an accumulation of analytical and numerical studies \cite{Bizon:2011gg,Dias:2011ss,Dias:2012tq,Buchel:2012uh,Buchel:2013uba,Bizon:2013xha,Maliborski:2013jca,Baier:2013gsa,Jalmuzna:2013rwa,Basu:2012gg,Friedrich:2014raa,Bizon:2017yrh,Maliborski:2014rma,Balasubramanian:2014cja,daSilva:2014zva,Craps:2014vaa,Dimitrakopoulos:2014ada,Horowitz:2014hja,Bizon:2015pfa,Basu:2015efa,Green:2015dsa,Deppe:2015qsa,Evnin:2015wyi,Craps:2015iia,Evnin:2015gma,Menon:2015oda,Freivogel:2015wib,Dias:2016ewl,Martinon:2017uyo,Dias:2017tjg,Choptuik:2017cyd,Choptuik:2018ptp}, is that islands of stability consist of initial data that is single-mode dominated, and data that is strongly multi-mode inevitably lead to horizon formation.

By providing a means for perturbation theory to break down, the resonant spectrum plays an important role in the instability of AdS. Indeed, the analytical studies suggest that a resonant spectrum is a necessary condition for an AdS instability \cite{Dias:2012tq,Menon:2015oda}.  See also \cite{Maliborski:2012gx,Okawa:2013jba,Okawa:2014nea,Okawa:2015xma,Maliborski:2014rma,Ponglertsakul:2016wae} for related numerical studies. Therefore, it is natural to expect that in cases where the spectrum is non-resonant, the islands of stability can contain strongly multi-mode data.

Dynamical evolution in AdS requires a choice of boundary conditions at the asymptotic boundary, and whether or not the spectrum is resonant depends on this choice.  Indeed, the resonant criterion requires a special choice of boundary conditions, which implies that the AdS instability requires a delicate fine-tuning of boundary conditions.

In this work, we investigate the affect of a resonant spectrum on the space of non-collapsing solutions.  We consider a massive complex scalar field $\psi$ in four dimensions, minimally coupled to gravity.  We choose a mass in AdS length units of $m^2=-2$ (we here and henceforth set the AdS length $\ell_{AdS}=1$).  This mass lies between the Breitenlohner-Freedman bound $m_{BF}^2=-9/4$ and the unitary bound $m_{*}^2=-5/4$. For this mass, Robin boundary conditions for the scalar field give normalizable modes, and correspond to a double-trace deformation of the boundary CFT \cite{Witten:2001ua}.  In general, the Robin boundary condition breaks the resonant spectrum of AdS, and is only asymptotically resonant.  However, the Robin condition can be made arbitrarily close to the Dirichlet or Neumann cases for which the spectrum is fully resonant.

With these boundary conditions, we will construct `multi-oscillators' \cite{Choptuik:2018ptp,Choptuik:2019zji}, which are solutions that are quasi-periodic in time, and hence do not collapse on any time scale. For simplicity, we will be chiefly concerned with two-mode, equal-amplitude data in spherical symmetry.  In the massless, Dirichlet case (which has a resonant spectrum), two-mode initial data is well-studied, and accumulated evidence suggests that such data always leads to collapse, and thus lies outside the islands of stability \cite{Dias:2012tq,Buchel:2012uh,Buchel:2013uba,Maliborski:2013jca,Green:2015dsa,Balasubramanian:2015uua,Bizon:2014bya,Balasubramanian:2014cja,Buchel:2014xwa,Dimitrakopoulos:2015pwa,Choptuik:2017cyd,Biasi:2018eaa,Biasi:2018isn,Choptuik:2018ptp,Biasi:2019bgq}.

Since multi-oscillators are quasi-periodic in time, they do not form black holes, and hence lie within the islands of stability.  In \cite{Choptuik:2018ptp}, these types of solutions were used in the massless Dirichlet case to map out portions of the island of stability.  Indeed, no equal-mode multi-oscillators were found, consistent with studies on two-mode initial data.

When the spectrum is non-resonant, as it is for Robin boundary conditions, analytic linear results suggest that any initial data sufficiently close to AdS should be stable.  One would therefore expect multi-oscillators to exist that are multi-mode dominated.  This is indeed what we find in this paper.  We also find that the space of these solutions appears to vanish as the boundary condition approaches the resonant Dirichlet case, in agreement with a non-linear instability for two-mode initial data in this case.  The range of existence of multi-oscillator solutions also allows us to quantify the extent to which breaking a resonant spectrum can restore the stability of AdS.

Our results in the Neumann case, however, were unexpected.  Though the spectrum is fully resonant, we nevertheless find non-collapsing strongly multi-mode initial data that are arbitrarily close to AdS. We find that these initial data admit a regular spectral decomposition.  However, we establish that these solutions are not well-approximated by two-mode perturbation theory, and confirm their apparent stability with a time evolution.  This provides a counterexample to the general picture that islands of stability for resonant systems are single-mode dominated.

\textbf{Setup --}  The ansatz for our metric and scalar field is given by
\begin{subequations}
\begin{align}\label{ansatz}
\mathrm ds^2&=\frac{1}{\cos^2 x}\left(-f \delta^2 \mathrm dt^2+\frac{dx^2}{f}+\sin^2x\,\mathrm d\Omega_{2}\right)\;,\\
\psi&=\cos x\,\phi\;,
\end{align}
\end{subequations}
where we take $f$, $\delta$, $\phi$ to be functions of $t$ and $x$, with $\phi$ complex. At the origin $x=0$, we require all fields to be regular. At $x=\pi/2$ we require the metric to asymptote to AdS, and choose the gauge where
\begin{align}
\delta\left(t,\frac{\pi}{2}\right)=1,\quad f\left(t,\frac{\pi}{2}\right)=1.
\end{align}
As we have mentioned in the introduction, we have to set the mass of the scalar field to be $m^2=-2$. With this mass, the scalar field has the boundary behaviour
\begin{align}
\phi(t,x)|_{x\to\frac{\pi}{2}}=\phi_1(t)+\phi_2(t)\left(x-\frac{\pi}{2}\right)+\mathcal O\left(x-\frac{\pi}{2}\right)^2.
\end{align}
We will consider for the scalar field, boundary conditions of the form
\begin{align}\label{boundaryconditions}
\sin\left(\frac{\pi}{2}\kappa\right)\phi_1\left(t\right)-\cos\left(\frac{\pi}{2}\kappa\right)\phi_2\left(t\right)=0,
\end{align}
with $0\leq\kappa\leq 1$.

The equations of motion in our setup consist of two spatial equations for $f$ and $\delta$, as well as a complex Klein Gordon equation for $\phi$.  There is additionally the Hamiltonian constraint equation (a temporal equation for $f$) which we do not solve directly, but use as a numerical consistency check.

When performing numerics, we will use the real functions $f_i,\ i\in\{1,2,3,4\}$ defined by
\begin{align}
\begin{split}
&\delta=1-\cos^2x\,f_1,\\
&f=1+\sin ^2x \cos ^{2}x\left(\cos x f_2+f_3^2+f_4^2\right),\\
&\phi=e^{i\omega_1 t}(f_3+if_4)\;,
\end{split}
\end{align}
where $\omega_1$ is a parameter that will be explained later.

In these variables, the Hamiltonian constraint equation implies that the following energy quantity is conserved in time:
\begin{align}\label{energy}
\mathcal E={f_2}\left(t,\frac{\pi}{2}\right)+\tan\left(\frac{\pi}{2}\kappa\right)\left({f_3}\left(t,\frac{\pi}{2}\right)^2+{f_4}\left(t,\frac{\pi}{2}\right)^2\right).
\end{align}

\textbf{Perturbative analysis --}
The Klein Gordon equation in an AdS background is a Sturm-Liouville problem with operator $L$:
\begin{align}
\partial_{t,t}\phi+L\phi=0,\quad L=-\frac{1}{\tan^2x}\partial_x\left(\tan^2x\partial_x\right).
\end{align}
This operator defines an inner product:
\begin{align}\label{scalarproduct}
(f,g):=\int_0^{\frac{\pi}{2}}f\,g\tan^2x\mathrm d x.
\end{align}
The Klein Gordon equation admits the following regular solutions:
\begin{align}\label{scalar}
\hat e_n(t,x)=C_n\,e^{i\omega_n t}\frac{\text{sinc}\left(\omega_n\, x\right)}{\text{sinc}\left(x\right)},
\end{align}
where the constants $C_n$ are chosen so that $(\hat e_n,\hat e_m)=\delta_{n,m}$. The boundary conditions \eqref{boundaryconditions} quantize the frequencies of the the scalar field \eqref{scalar}. These frequencies $\omega_n$ are given by the solutions to the equation:
\begin{align}
\sin\left(\frac{\pi}{2}\kappa\right)\sin\left(\frac{\pi}{2}\omega_n\right)-\omega_n\cos\left(\frac{\pi}{2}\kappa\right)\cos\left(\frac{\pi}{2}\omega_n\right)=0.
\end{align}

Only two values of $\kappa$ lead to frequencies that are evenly spaced.  A Dirichlet boundary condition corresponds to $\kappa=1$ with frequencies $\omega_n=2n$, and a Neumann boundary condition corresponding to $\kappa=0$ with frequencies $\omega_n=2n+1$.  These evenly spaced modes create a fully resonant spectrum that will cause secular terms to appear at higher orders in perturbation theory.  In particular, when the spectrum of frequencies is such that there exist quadruples of frequencies $\{j_1,j_2,j_3,j_4\}$ obeying
\begin{align}\label{resonance}
\Delta\omega_J\equiv \omega_{j_2}+\omega_{j_3}-\omega_{j_4}-\omega_{j_1}=0,
\end{align}
then secular terms appear at higher order in a perturbative analysis.  These secular terms cannot be removed by shifts in the frequencies, and are absent only when starting with a single mode at the lowest order in perturbation theory.  These secular terms therefore lead to a breakdown of perturbation theory for multi-mode data.

These frequencies are continuously connected by varying $\kappa$.  However, note we only take positive definite frequencies, so $\omega_0$ does not exist for the Dirichlet boundary condition, but does exist for the Neumann condition. By continuity, this frequency $\omega_0$ appears when $\kappa\leq\frac{2}{\pi } \tan ^{-1}\left(\frac{2}{\pi }\right)$.

\textbf{Double Oscillators --}  We now study the space of non-collapsing solutions by constructing double-oscillators, which are quasi-periodic solutions that oscillate on two frequencies \cite{Choptuik:2018ptp,Choptuik:2019zji}.  One of the frequencies is given by the parameter $\omega_1$ as we have defined in the ansatz \eqref{ansatz}. The other frequency is obtained by demanding that the functions be periodic in time with period $\omega_2$.  Specifically, we require that the functions take the Fourier expansion
\begin{align}
    f_i(t,x)&=\sum_k f_{i}^{(k)}(x)\cos(k\omega_2 t)\qquad i\in\{1,2,3,\}\;,\nonumber\\
    f_4(t,x)&=\sum_k f_{4}^{(k)}(x)\sin(k\omega_2 t)
\end{align}
for Fourier coefficients $f_i^{(k)}$.

Because of the quasi-periodicity, double-oscillators do not collapse to form black holes, and hence lie within the islands of stability.  In fact, double-oscillators can be extended to form the more general multi-oscillator family by including more frequencies in their quasiperiodicity. The multi-oscillator family therefore has an infinite number of parameters, and their existence can be used to assess the size of the islands of stability.

The equations of motion are then solved as a boundary value problem with periodic boundary conditions in time. Numerically, we use a Newton-Raphson method with the perturbative solution as an initial estimate. We utilize pseudo-spectral discretization with a half-Fourier grid in the time direction and Legendre-Gauss-Lobatto nodes for the spatial direction.

The solutions can be parametrised by $\kappa$ and the frequencies $\omega_1$ and $\omega_2$. For our purposes, it is more convenient to use the projection onto the normal mode functions \eqref{scalar} under the inner product \eqref{scalarproduct}.  We consider two-mode-equal-amplitude solutions which satisfy
\begin{align}\label{twomodeamplitude}
(f_{3}^{(1)},\hat e_{n_1})=(f_{4}^{(2)},\hat e_{n_2})=\varepsilon\;,
\end{align}
and parametrise our solutions by $\varepsilon$ and $\kappa$. We will consider the case where $n_1=1$ and $n_2=2$. Numerically, the integrals involved in the projections \eqref{twomodeamplitude} are performed by Gaussian quadrature.

Our first double-oscillator solutions were found for intermediate values of $\kappa$ with small values of $\varepsilon$ (and consequently also small energy $\mathcal E$ as defined in in \eqref{energy}).  The existence of these solutions for intermediate values of $\kappa$ with arbitrarily small energy is consistent with the expectation of stability for these boundary conditions.

As $\varepsilon$ is increased, $\mathcal E$ also increases until we eventually cease to find solutions.  The point at which solutions cannot be found appears to be robust to changes in grid size and parametrization, and can be located using a bisection search.

\begin{figure}
\centering
\includegraphics[width=.4\textwidth]{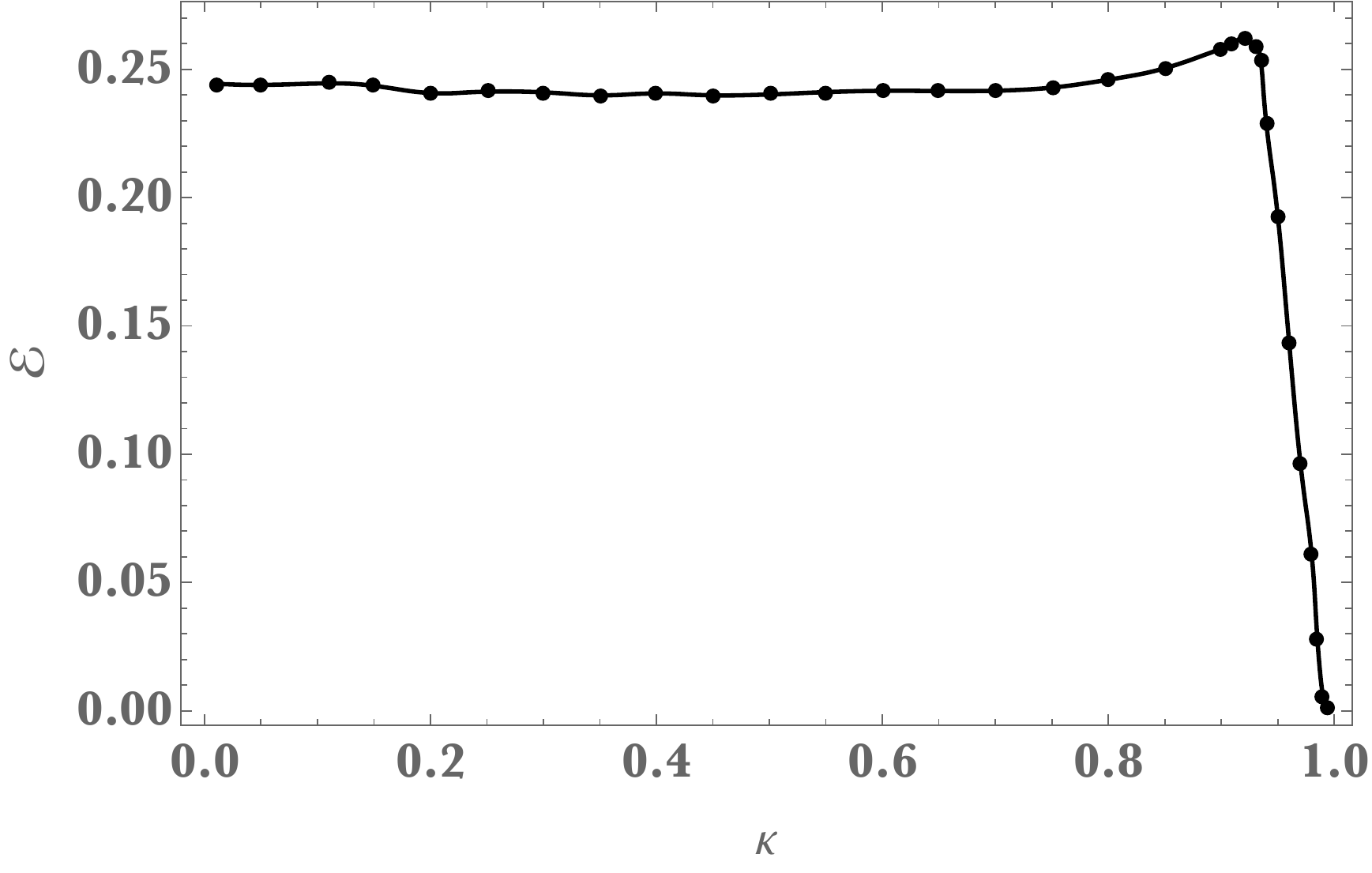}
\caption{Maximum value of the energy for which we have encountered mulit-oscillators as a function of $\kappa$.  }\label{fig:maxenergy}
\end{figure}

In figure \ref{fig:maxenergy} we display the maximum value of $\mathcal E$ for which we have obtained double-oscillator solutions as a function of the parameter $\kappa$.  Observe that this curve approaches zero near the Dirichlet case ($\kappa=1$).  This is in agreement with the expectation that two-mode-equal-amplitude data will eventually form black holes in the Dirichlet case, which implies that double-oscillators cannot exist.

However, this curve does not vanish on the Neumann case $\kappa=0$, and instead plateaus to a constant value.  Even though the Neumann case has a resonant spectrum that predicts that perturbation theory breaks down for multi-mode data, the islands of stability still contain strongly multi-mode data.

These results point at the existence of islands of stability for Neumann boundary conditions that do not have a counterpart in the Dirichlet case.  In the Dirichlet case for a massless scalar, the emerging picture has been that islands of stability are single-mode dominated.  In fact, in \cite{Choptuik:2018ptp} the authors charted islands of stability by constructing double-oscillators that branch from boson stars, and found that all such solutions remained single-mode dominated. In the present Neumann case, however, the double-oscillators are multi-mode dominated.

\begin{figure}
\centering
\includegraphics[width=.4\textwidth]{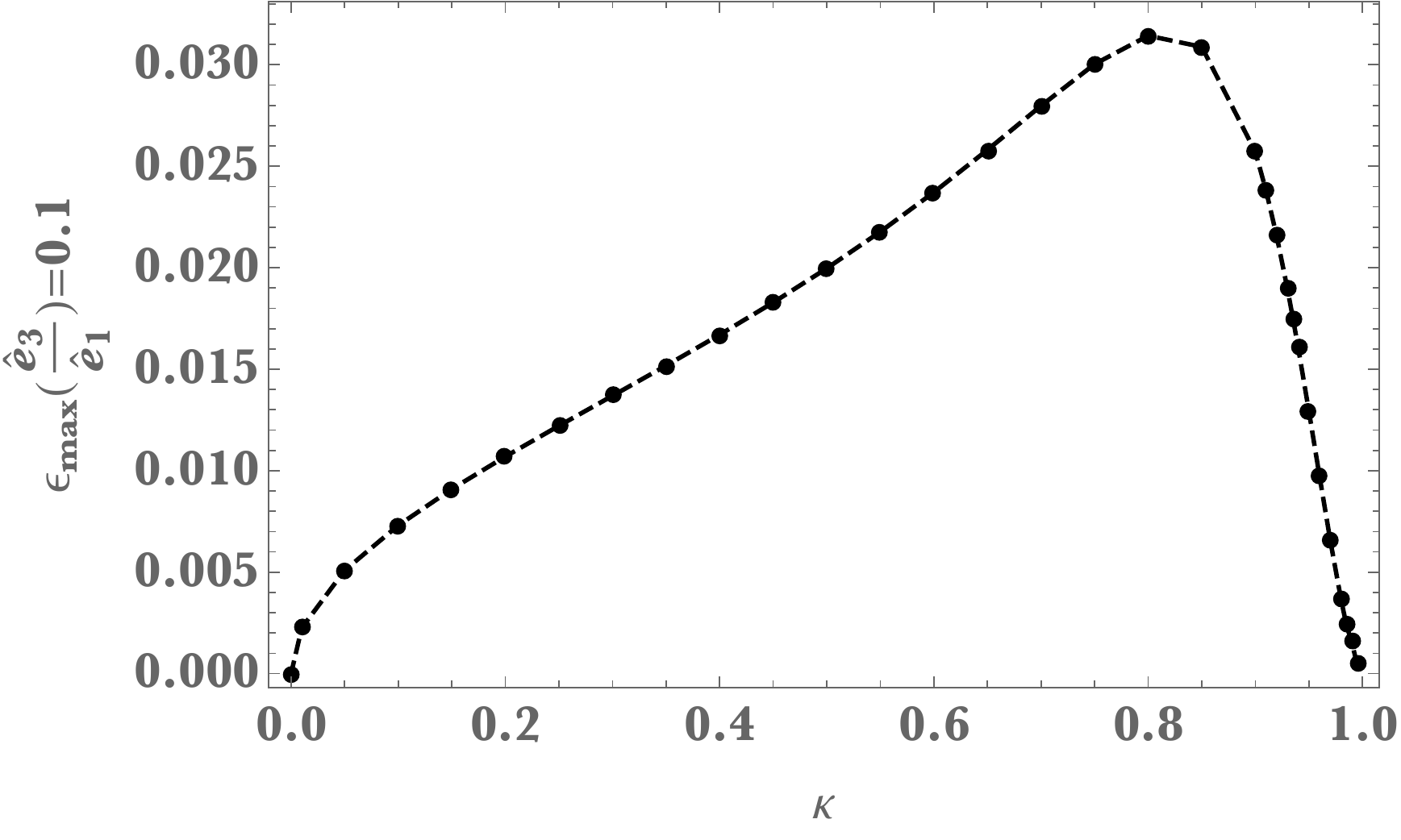}
\caption{As a function of $\kappa$, the maximum value of the energy for which we have encountered multi-oscillators where the ratio of the amplitudes of the projections to $\hat e_1$ and $\hat e_3$ remains smaller than 10\%.   }\label{fig:thirdmode}
\end{figure}

Let us now attempt to reconcile the existence of strongly multi-mode oscillators with the breakdown of perturbation theory due to resonant secular terms.  Specifically, we will verify that double oscillators for near-Neumann boundary conditions are not described by perturbation theory with two-mode data.  Perturbation theory for strictly two-mode data requires that for small energies, the double oscillators should approach the solution $\phi=\varepsilon( \hat e_{n_1}+\hat e_{n_2})$.  In particular $(\phi,\hat e_{n_3})/(\phi,\hat e_{n_1})$ must vanish at small $\varepsilon$ for any third mode $n_3$.  In figure \ref{fig:thirdmode} we display the maximum amplitude for which $(f_{3}^{(3)},\hat e_{n_3})/(f_{3}^{(1)},\hat e_{n_1})<0.1$ in our double oscillator solutions. Notice that this curve vanishes near the Neumann boundary condition.  Here, we have chosen $n_3=3$, but the curve looks qualitatively similar for other $n_3$.  In other words, the higher modes remain large relative to the lower modes in the Neumann case, even for small energies.

\begin{figure}
\centering
\includegraphics[width=.4\textwidth]{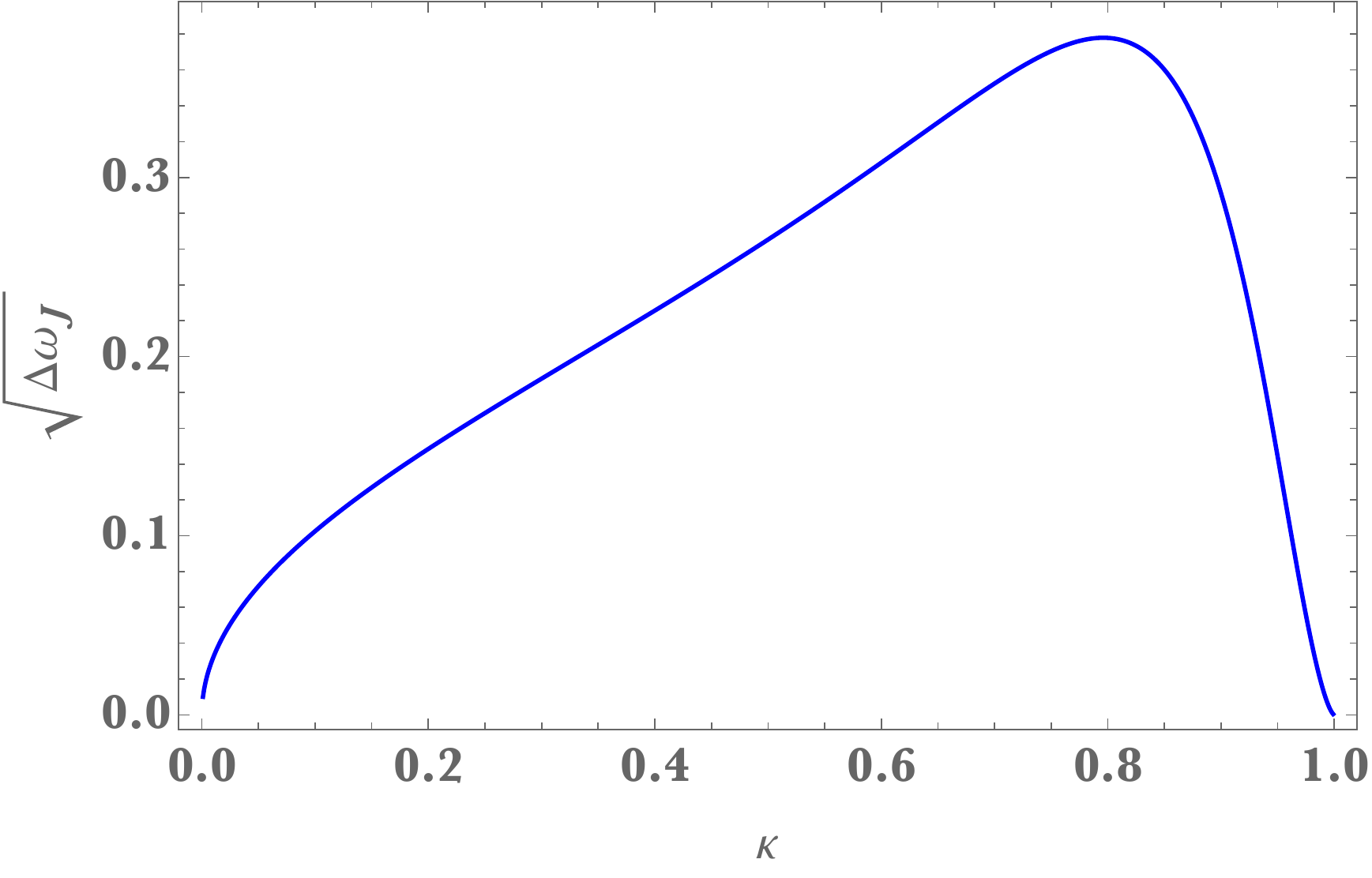}
\caption{Value of $\sqrt{\Delta\omega_J}$ as a function of $\kappa$. $\kappa=0$ is the Neumann boundary condition on the scalar field and $\kappa=1$ a Dirichlet boundary condition. In both these cases, $\Delta\omega_J=0$ and the spectrum is fully resonant.}\label{fig:resonance}
\end{figure}

We can compare this behaviour to $\sqrt{\Delta\omega_J}$, defined in \eqref{resonance}, which can be used a measure of how close we are to a resonant spectrum. Perturbation theory shows that the frequency corrections appear at order $\varepsilon^2$. Hence it is natural to use the square root when comparing frequencies with amplitudes.   We display this quantity as a function of $\kappa$ in Fig.~\ref{fig:resonance}.  We can see that both \ref{fig:thirdmode} and Fig.~\ref{fig:resonance} are qualitatively similar.

\begin{figure}
\centering
\includegraphics[width=.4\textwidth]{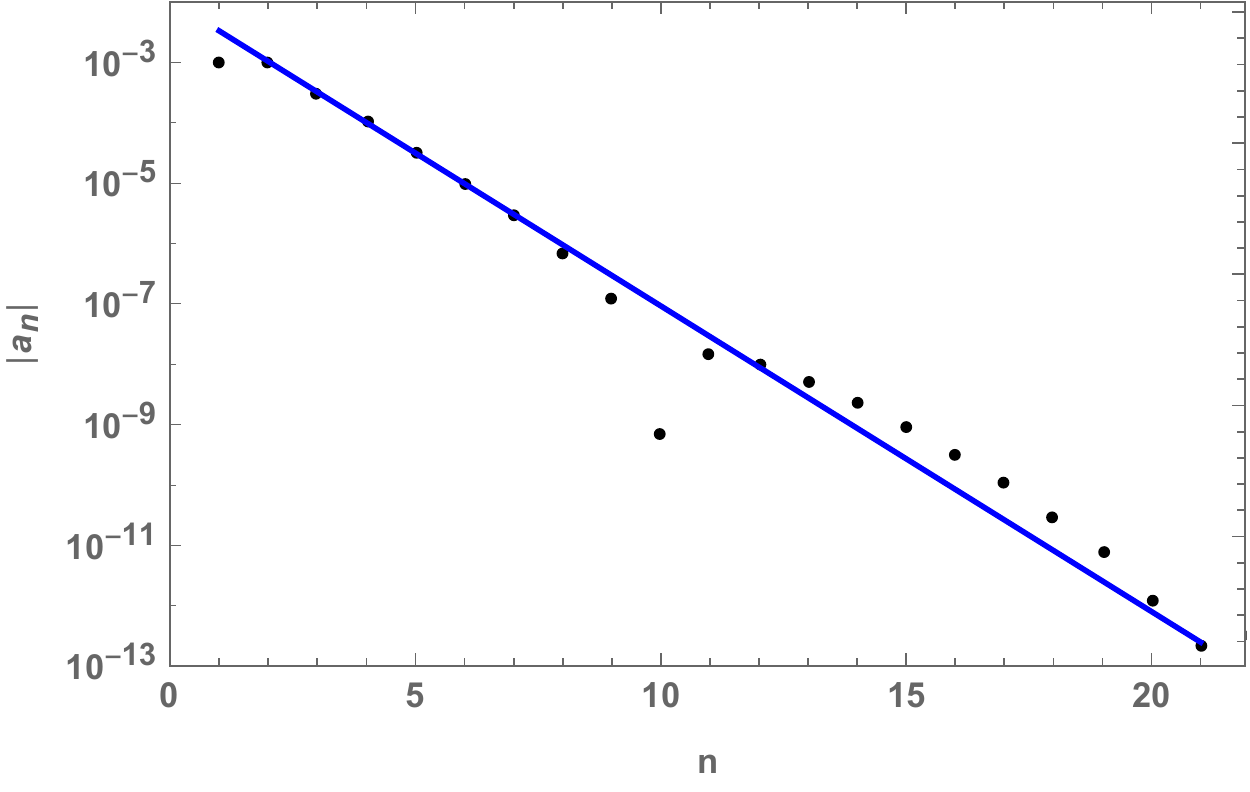}
\caption{Mode amplitudes $|a_n|$ of double oscillators for Neumann boundary conditions ($\kappa=0$) and $\varepsilon=0.001$. The amplitude is obtained by projecting the time slice $t=0$ onto the normal modes.}\label{fig:spectrum}
\end{figure}

We therefore see that at small energies, the Neumann double-oscillators do not approach two-mode data since a third mode $n_3$ remains large. We now study the fall-off in amplitude of these higher modes as a function of the energy.  In order to obtain the spectrum we considered the time slice $t=0$ and projected onto the normal modes:
\begin{equation}
a_n\equiv(f_3(t=0,x),\hat e_n)
\end{equation}
In figure \ref{fig:spectrum} we show the particular spectrum $|a_n|$ of double-oscillators with Neumann boundary conditions and $\varepsilon=0.001$. We see that the first two amplitudes of $|a_n|$ are equal, as expected by our equal-mode boundary conditions. For higher modes, there is an exponential fall-off, as is typical for smooth solutions.  However, for very high mode numbers (higher than is shown), there is eventually a power-law tail due to the fact that the normal mode basis functions are even about the AdS boundary, while the fully non-linear solutions are not; they merely satisfy a Neumann condition, and are still smooth solutions.  The smoothness of double-oscillators can be seen in a Fourier-Legendre basis, where the spectrum has exponential fall-offs and no power-law tail.

\begin{figure}
\centering
\includegraphics[width=.4\textwidth]{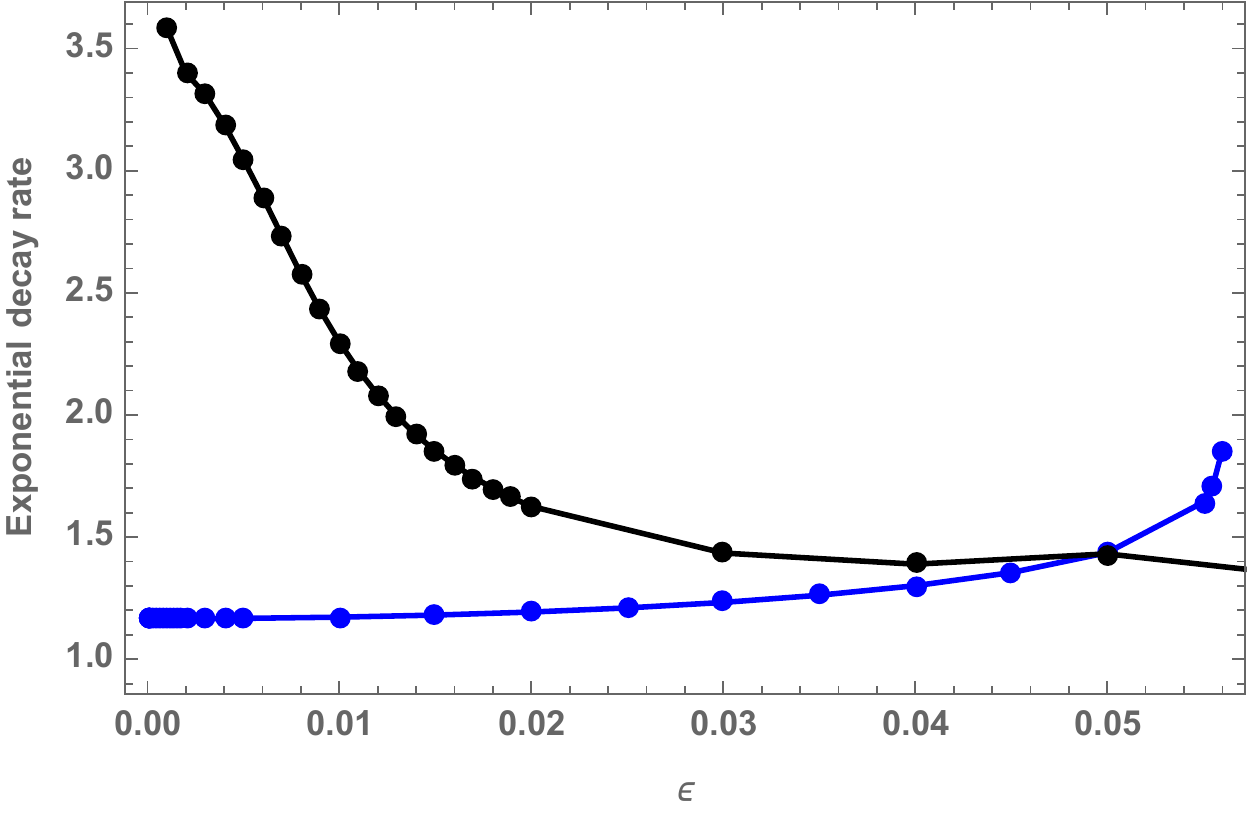}
\caption{Decay rate of mode amplitudes as a function of $\varepsilon$. The decay rate is obtained by projecting double-oscillators onto the normal modes, followed by a fit using linear regression. In black we show double-oscillators correspond to Robin boundary conditions with $\kappa=0.15$ whereas in blue we show the Neumann case.}\label{fig:slope}
\end{figure}

We are interested in the initial exponential fall-off of $|a_n|$, and not the later power-law tail. In figure \ref{fig:slope}, we show the exponential decay of normal mode amplitudes as a function of $\varepsilon$ for Neumann boundary conditions (blue) and a $\kappa=0.15$ Robin boundary condition (black). In the Robin case, the modes decay more rapidly as we approach the perturbative limit $\varepsilon\to 0$. This is the expected result since in the perturbative limit we find two modes with amplitude $\varepsilon$ while other modes are highly suppressed. We encounter the same qualitative behaviour for other values of $\kappa$ different from 0. In the Neumann case, however, notice that the decay rate approaches a constant value as $\varepsilon\to 0$. This indicates that this limit does not resemble two-mode data but rather multi-mode data.

Therefore, to access the Neumann double-oscillator solution perturbatively, one would need to seed the perturbation expansion with an infinite number of modes.  It is conceivable that in such a situation, the secular terms can become supressed.  We note that as $\varepsilon\to 0$, the power-law tail gets pushed to higher and higher modes, suggesting that the perturbative solution may be an even function about the AdS boundary.

\textbf{Numerical evolution --}
Though multi-oscillator solutions never collapse to form black holes, they can be unstable.  Indeed, solutions such as boson stars \cite{Dias:2011at,Liebling:2012fv,Buchel:2013uba,Choptuik:2017cyd} and oscillatons \cite{Maliborski:2013jca,Fodor:2015eia} form a subset of the multi-oscillator family and are known to be unstable to black hole formation for sufficiently large energies.  In a sense, these unstable solutions would constitute regions of the islands of stability that are locally measure zero.

We therefore assess the stability of double-oscillators in the Neumann case through a numerical time evolution. As initial data, we provide the double-oscillator solutions at time $t=0$ and let them evolve. We performed the study for different values of $\varepsilon$. In the following we consider the study with $\varepsilon=0.06$ which corresponds to energy $\mathcal E=0.24$.

In the AdS instability, gravitational collapse typically occurs on a timescale of order $t\sim\varepsilon^{-2}$ \cite{Bizon:2011gg}, which happens to be the fastest timescale allowed by the breakdown of perturbation theory.  We let our code evolve until time $t=\varepsilon^{-3}\approx 4630$. As in the boundary value problem, we controlled our numerics through energy conservation and the constraint equation. Both the constraint equation and energy conservation were verified to order $10^{-10}$.

\begin{figure}
\centering
\includegraphics[width=.4\textwidth]{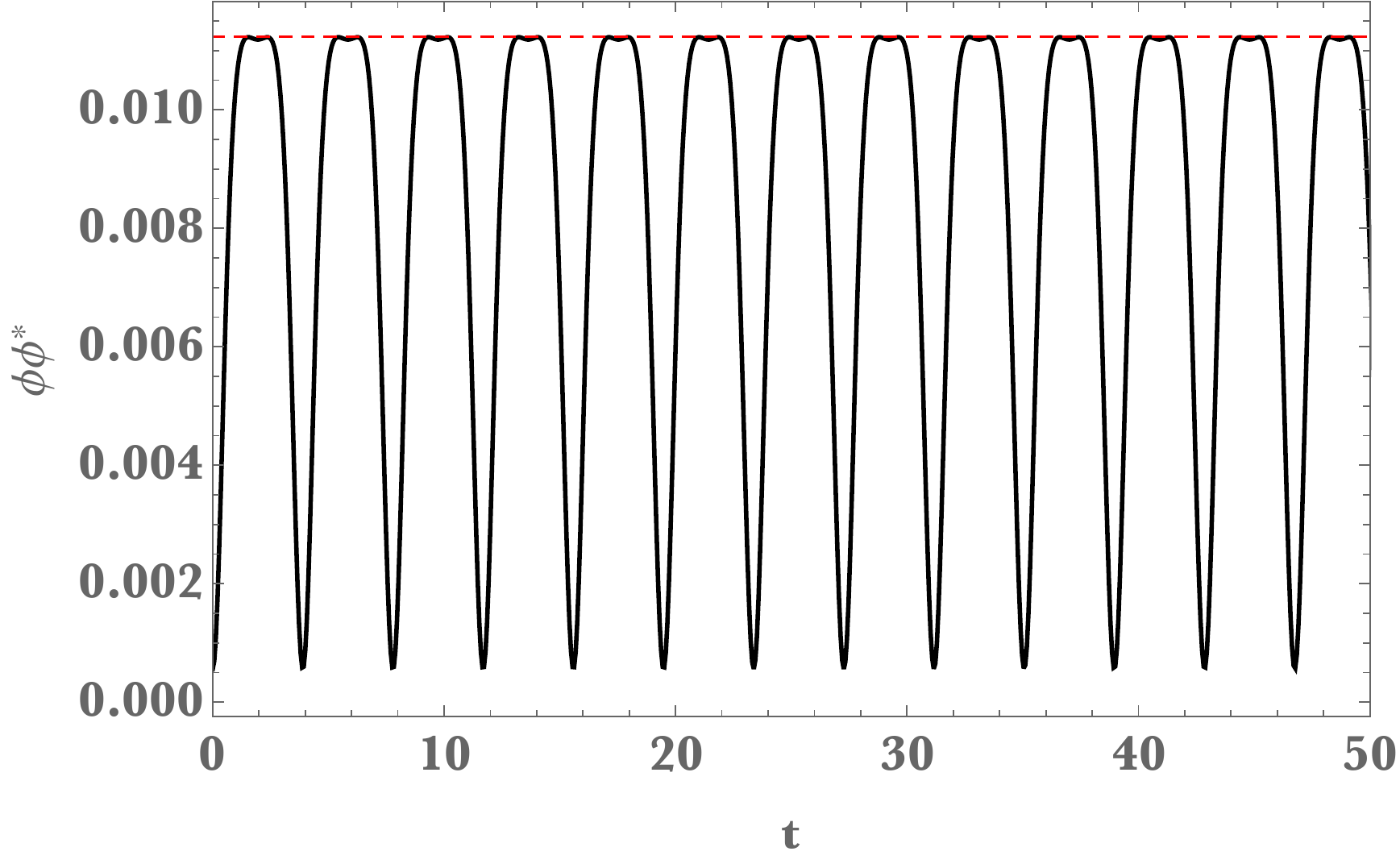}
\includegraphics[width=.4\textwidth]{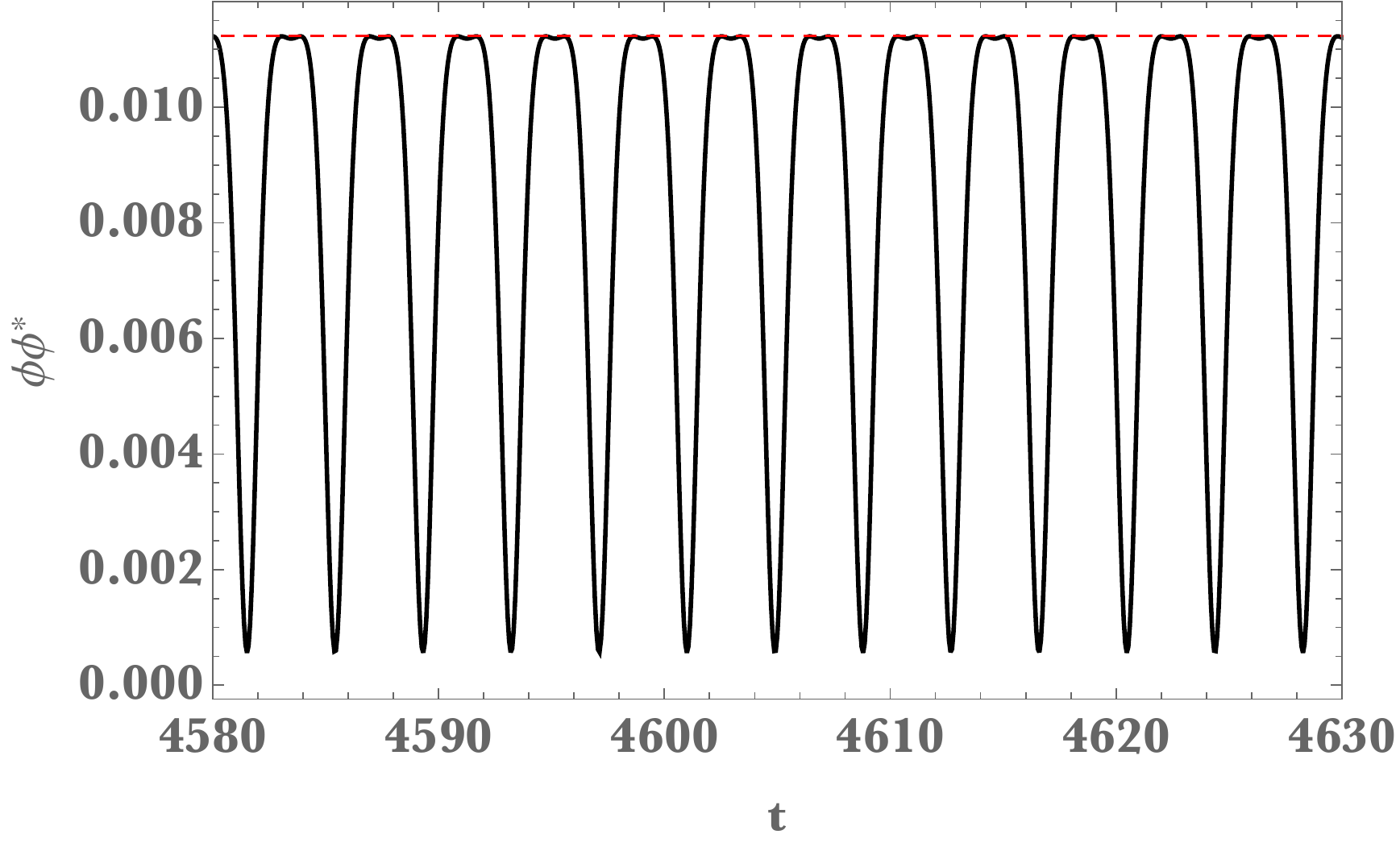}
\caption{ $\phi\phi^*$  at the boundary $x=\pi/2$ as a function of time for initial data consisting of a multi-oscillator with two main amplitudes $\varepsilon=0.06$. {\emph{Top}}: evolution at early times. {\emph{Bottom}}: evolution at late time.}\label{fig:timeevol}
\end{figure}

The multi-oscillator solutions oscillate in two frequencies. Thus the gauge invariant quantity $\phi\phi^*$ should be periodic as the product cancels one of the frequencies. We thus plot the quantity $\phi\phi^*$ at the boundary as a function of time in figure \ref{fig:timeevol}. The panel above corresponds early-time evolution, whereas the panel below shows late time. Clearly, the solutions are non-collapsing and also remain periodic throughout the time of simulation.  This provides evidence that these double-oscillators in the Neumann case are nonlinearly stable.

\textbf{Discussion --}
To summarise, we have studied the existence of two-mode-equal-amplitude double-oscillator solutions in asymptotically AdS spacetimes with Robin boundary conditions that generically break the resonant spectrum.  These boundary conditions enables us to move away from the resonant case of Dirichlet boundary conditions ($\kappa=1$) up to Neumann boundary conditions ($\kappa=0$) which is also resonant.

The existence of these double-oscillators when the spectrum is non-resonant shows that islands of stability in these cases can contain strongly multi-mode data.  This result is consistent with the idea that AdS is stable when the spectrum is non-resonant.  The fact that double-oscillators vanish as one approaches the Dirichlet case is consistent with the idea that two-mode data form black holes.

But surprisingly, we have found stable (until at least $t\sim1/\epsilon^2$), strongly multi-mode data in the Neumann case, even though the linear spectrum is resonant.  While a fully resonant spectrum might be necessary for a nonlinear instability, it is not sufficient.  To our knowledge, this is the first example in AdS of a non-collapsing strongly multi-mode data in a resonant system.

We have performed analytical perturbative studies, similar to \cite{Bizon:2011gg}, in order to study and compare the differences between Dirichlet and Neumann boundary conditions.  In general, the secular terms at $O(\epsilon^3)$ coming from two-mode data at $O(\epsilon)$ can actually be removed by turning on a third mode at $O(\epsilon)$.  However, the introduction of this third mode causes other secular terms to appear.  Those can again be removed by including more modes at $O(\epsilon)$, but always at the cost of other secular terms appearing.

We have considered two, three and four mode initial data with the amplitude of the two lowest modes fixed to be the same, and tuning the additional modes to cancel secular terms that have already appeared. In the Neumann case, we find that the additional secular terms that appear have smaller and smaller amplitude as more modes are added at $O(\epsilon)$.  This seems to hint that the secular terms may become more and more negligible as more modes are added at $O(\epsilon)$, perhaps vanishing in some limit. This is to be contrasted to the Dirichlet case, where these extra secular terms appear with similar amplitude.

Rather than continuing in this manner in perturbation theory, the two-time-formalism \cite{Balasubramanian:2014cja,Craps:2014vaa,Buchel:2014xwa,Bizon:2015pfa,Evnin:2015wyi,Biasi:2018eaa,Biasi:2018isn,Biasi:2019bgq} provides an alternative description at $O(\epsilon^3)$ that does not have resonances, and might be able to shed light on the origin of these solutions.  This investigation lies beyond the scope of the present study and we leave it for future work.

In this paper, we have shown results for the modes $\hat e_1$ and $\hat e_2$ in \eqref{scalar}. We have also performed all the studies for $\hat e_0$ and $\hat e_1$ for values of $\kappa$ where $\hat e_0$ exists. Our results are qualitatively similar.

We have also repeated the study in this paper in the flat-space case of a Dirichlet box \cite{Maliborski:2012gx}. There we have considered a massive scalar field where the mass breaks the resonant spectrum, and therefore plays the role of the parameter $\kappa$. In that case, the spectrum is resonant only when the scalar is massless, and non-resonant otherwise.  The behaviour for small mass is qualitatively similar to what was found in AdS for $\kappa\approx 1$. For the flat-space Dirichlet box, we were unable to find behaviour similar to that of a scalar field in AdS with Neumann boundary conditions.

The existence of stable strongly-multi-mode data with Neumann boundary conditions leaves open the question of horizon formation for some other set of arbitrarily small energy data.  Initial data that is two-mode dominated (with all the higher modes greatly supressed) lies far away from the double-oscillators we found, and might still be nonlinearly unstable to forming black holes.  It would be interesting to explore this family of two-mode initial data, investigate its propensity to black hole formation, and understand its relationship to the double-oscillators we have constructed.


{\bf~Acknowledgements --} We thank Oscar Dias, Gary Horowitz, Andrzej Rostworowski, Jorge Santos, and David Turton for helpful comments. R.M. acknowledges support from STFC Ernest Rutherford grant ST/M004147/1 and Univ. Southampton Global Partnerships Award 2018-19. B.W. is supported by NSERC. R.M. would like to thank the University of British Columbia for hospitality during the completion of this work.
\bibliography{refsmultiosc}{}
\end{document}